\documentstyle[aps,eqsecnum,floats]{revtex}
\begin{document}
\draft
\twocolumn[\hsize\textwidth\columnwidth\hsize\csname
           @twocolumnfalse\endcsname
\title{Bounding the mass of the graviton using gravitational-wave\\
       observations of inspiralling compact binaries}
\author{Clifford M.~Will \cite{cmw}}
\address{McDonnell Center for the Space Sciences, 
         Department of Physics, Washington University,
         St.~Louis, Missouri 63130}
\maketitle
\begin{abstract}
\widetext
If gravitation is propagated by a massive field, then the velocity of
gravitational waves (gravitons) 
will depend upon their frequency as $(v_g/c)^2=1-(c/f\lambda_g)^{2}$, and 
the effective
Newtonian potential will have a Yukawa form $\propto
r^{-1} \exp(-r/\lambda_g)$, where $\lambda_g=h/m_g c$ is the graviton Compton
wavelength.  In the case of
inspiralling compact binaries, gravitational waves emitted at low frequency
early in the inspiral will travel slightly slower than those emitted
at high frequency later, resulting in an offset in the relative
arrival
times at a detector.  This modifies the phase evolution of the
observed inspiral gravitational waveform, similar to that caused by
post-Newtonian corrections to quadrupole phasing.  Matched filtering
of the waveforms can bound such frequency-dependent variations
in propagation speed, and thereby bound the graviton
mass.  The bound depends on the mass of the source and on noise
characteristics of the detector, but is independent of the distance to
the source, except for weak cosmological redshift effects.  
For observations of stellar-mass compact inspiral using
ground-based interferometers of the LIGO/VIRGO type, 
the bound on $\lambda_g$ is of the order of $6 \times 10^{12}$ km, 
about double
that from solar-system tests of Yukawa modifications of Newtonian
gravity.  For observations of super-massive black hole binary inspiral
at cosmological distances
using the proposed laser interferometer space antenna (LISA), 
the bound can be as large as $6 \times 10^{16}$ km.  This is three
orders of magnitude weaker than model-dependent bounds from galactic
cluster dynamics.
\end{abstract}
\vskip 2pc]
\narrowtext

\section{Introduction}
The detection of gravitational radiation by either laser
interferometers or resonant cryogenic bars will, it is widely stated,
usher in a new era of gravitational-wave astronomy \cite{Thorne1987}.  
Furthermore,
according to conventional wisdom, it
will yield new and interesting tests of general relativity (GR) in its
radiative regime.  These tests are generally based on three aspects
of gravitational radiation: its back-reaction on the source, its 
polarization, and
its speed. 

{\it (i) Gravitational back-reaction.} \quad 
This plays an important role only in the
inspiral of compact objects.  The equations of motion of inspiral
include the non-radiative, non-linear post-Newtonian corrections
of Newtonian motion,
as well as radiation back-reaction and its non-linear post-Newtonian
corrections.  The evolution of the orbit is imprinted on the phasing
of the inspiral waveform, to which broad-band laser interferometers
are especially sensitive through the use of matched filtering of the
data against theoretical templates derived from GR.  A
number of tests of GR  using matched filtering of binary inspiral
have been discussed, including putting a bound on scalar-tensor
gravity \cite{willbd}, measuring the non-linear ``tail term'' in
gravitational radiation damping \cite{lucsathya}, and testing the GR
``no hair'' theorems by mapping spacetime outside 
black holes \cite{ryan,eric}.  A concrete
test of gravitational back-reaction, albeit
at the lowest order of approximation, 
has already been provided by the Binary
Pulsar PSR 1913+16, where the tracer of the orbital phase was the radio
emission from a
pulsar rather than matched filtering of gravitational waves \cite{taylor}.

{\it (ii) Polarization of gravitational waves.} \quad 
In GR, gravitational waves come in at most two polarization states,
independently of the source, while in alternative theories of gravity,
there are as many as six polarizations
\cite{polarize,tegp}.
Using a sufficiently large number of gravitational antennas suitably
oriented, it is possible to determine or limit the polarization
content of an incident wave, and thereby to test theories.  For
example, should an incident wave be shown definitively to have three
polarizations, the result would be devastating for GR.  Although some
of the details of implementing such polarization observations have
been worked out for arrays of resonant cylindrical, disk-shaped, and
spherical detectors 
\cite{polarize,paik}, rather little has been done to assess whether the
ground-based
laser-interferometers (LIGO, VIRGO, GEO600, TAMA) could perform
interesting polarization measurements.  The results depend
sensitively on the relative orientation of the detectors' arms, which are now
cast (literally) in concrete.

{\it (iii) Speed of gravitational waves.} \quad 
According to GR, in the limit in which the wavelength of gravitational
waves is small compared to the radius of curvature of the background
spacetime, the waves propagate along null geodesics of the background
spacetime, {\it i.e.} they have the same speed, $c$, as light.  In other
theories, the speed could differ from $c$ because of coupling of
gravitation to ``background'' gravitational fields.  For example, in
the Rosen bimetric theory \cite{rosen} with a flat background metric
$\mbox{\boldmath$\eta$}$, 
gravitational waves follow null geodesics of $\mbox{\boldmath$\eta$}$,
while light follows null geodesics of ${\bf g}$ \cite{caves,tegp}.  

Another way in which the speed of gravitational waves could differ
from one is if gravitation were propagated by a massive field (a
massive graviton), in which case, $v_g$ would
be given by, in a local inertial frame,
\begin{equation}
{v_g^2 \over c^2} = 1- {m_g^2c^4 \over E^2} \,,
\label{eq1}
\end{equation}
where $m_g$ and $E$ are the graviton rest mass and energy,
respectively.

The most obvious way to test this is to
compare the arrival times of a gravitational wave and an electromagnetic
wave from the same event, {\it e.g.} a supernova.
For a source at a distance $D$, the
resulting value of the difference $1-v_g/c$ is 
\begin{equation}
1- {v_g \over c}= 5 \times 10^{-17} \left ( 
{{200 {\rm Mpc}} \over D} \right ) \left ( {{\Delta t}
\over {1 {\rm s}}} \right )
\,,
\label{eq2}
\end{equation}
where 
$\Delta t$ is the ``time difference'', given by
\begin{equation}
\Delta t \equiv \Delta t_a - (1+Z) \Delta t_e  \,,
\label{eq3}
\end{equation}
where $\Delta t_a$ and $\Delta t_e$ are the differences 
in arrival time and emission time, respectively, of the
two signals, and $Z \simeq DH_0/c$ is the redshift of the source, with $H_0$
the Hubble parameter.  
In many cases, $\Delta t_e$ is unknown, 
so that the best one can do is employ an upper bound on
$\Delta t_e$ based on observation or modelling.
The result will then be a bound on $1-v_g/c$.  

If the frequency of the gravitational waves is such that $hf \gg m_gc^2$,
where $h$ is Planck's constant, then $v_g/c \approx 1- {1 \over 2} (c/\lambda_g
f)^{2}$, where $\lambda_g= h/m_gc$ is the graviton Compton wavelength, and
the bound on $1-v_g/c$ can be converted to a bound on $\lambda_g$, given
by
\begin{equation}
\lambda_g > 3 \times 10^{12}{\rm km} \left ( {D \over {200 {\rm
Mpc}}}  
{{100 {\rm Hz}} \over f} \right )^{1/2} \left ({1 \over
{f\Delta t}} \right )^{1/2} \,.
\label{eq4}
\end{equation}

The foregoing discussion assumes that the source emits {\it both}
gravitational and electromagnetic radiation in detectable amounts, and
that the relative time of emission can be established (by one means or
another) to sufficient accuracy, or can be shown to be sufficiently
small.

However, there is a situation in which a bound on the graviton mass
can be set using gravitational radiation alone.  That is the case of
the inspiralling compact binary.  Because the frequency of the
gravitational radiation sweeps from low frequency at the initial
moment of observation to higher frequency at the final moment, the
speed of the gravitons emitted will vary, from lower speeds initially
to higher speeds (closer to $c$) at the end.  This will cause a
distortion of the observed phasing of the waves and result in a
shorter than expected
overall time $\Delta t_a$ of passage of a given number of cycles.  
Furthermore, through the technique of matched filtering, the
parameters of the compact binary can be measured accurately
\cite{jugger}, and
thereby the emission time $\Delta t_e$ can be determined accurately.
Roughly speaking, the ``phase interval'' $f\Delta t$ in Eq. (\ref{eq4}) can be
measured to an accuracy $1/\rho$, where $\rho$ is the signal-\-to-\-noise
ratio.  

Thus we can estimate the bounds on $\lambda_g$ achievable for various
compact inspiral systems, and for various detectors.   For
stellar-mass inspiral (neutron stars or black holes) observed by the
LIGO/VIRGO class of ground-based interferometers, we have $D \approx
200 {\rm Mpc}$, $f \approx 100 {\rm Hz}$, and $f\Delta t \sim
\rho^{-1} \approx 1/10$ \cite{LIGO}.  
The result is $\lambda_g > 10^{13}
{\rm km}$.  For supermassive binary black holes ($10^4$ to $10^7
M_\odot$) observed by the proposed laser-interferometer space antenna
(LISA), we have $D \approx 3 {\rm Gpc}$, $f \approx 10^{-3} {\rm Hz}$,
and $f\Delta t \sim
\rho^{-1} \approx 1/1000$ \cite{LISA}.  The result is $\lambda_g >  10^{17}
{\rm km}$.

\begin{table}[t]
\caption{Bounds on $\lambda_g$ from gravitational-wave observations of
inspiralling compact binaries, using ground-based (LIGO/VIRGO) and
space-based (LISA) observatories.  Masses are in $M_\odot$. }
\begin{tabular}{cccc}
$m_1$&$m_2$&Distance (Mpc)&Bound on $\lambda_g$ (km)\\
\hline
\multicolumn{4}{l}{Ground-based (LIGO/VIRGO)} \\

1.4&1.4&300&$4.6 \times 10^{12}$\\
1.4&10&630&$5.4 \times 10^{12}$\\
10&10&1500&$6.0 \times 10^{12}$\\
\hline
\multicolumn{4}{l}{Space-based (LISA)} \\
$10^7$&$10^7$&3000&$6.9 \times 10^{16}$\\
$10^6$&$10^6$&3000&$5.4 \times 10^{16}$\\
$10^5$&$10^5$&3000&$2.3 \times 10^{16}$\\ 
$10^4$&$10^4$&3000&$0.7 \times 10^{16}$\\
\end{tabular}
\label{summary}
\end{table}

We have refined these crude estimates by explicit calculations using matched
filtering (Table \ref{summary}).  We first calculate the effect of the 
frequency-dependent massive graviton velocity on 
the observed gravitational-wave
phasing.  We assume that the evolution of the system, driven by 
gravitational back-reaction,
is given correctly by general relativity,
apart from corrections of fractional order 
$(r/\lambda_g)^2$, where $r$
is the size of the binary system; these corrections can be shown to be
negligible for the cases of interest.  Including GR post-Newtonian (PN) 
and tail terms (1.5PN) in the phasing, and
assuming circular orbits and non-spinning bodies, we determine the
accuracy with which the parameters of the system can be measured
(``chirp'' mass of the system, reduced mass, fiducial phase, and 
fiducial time), and simultaneously find the
accuracy with which the effect of a graviton mass can be
bounded (effectively, we find an upper bound on
$\lambda_g^{-1}$).  We use noise curves appropriate for the advanced
LIGO detectors, and for the proposed LISA observatory.  It is
interesting to note that, despite the apparent distance dependence in
Eq. (\ref{eq4}), the bound for a given system is independent of its distance,
because the signal-to-noise ratio, which determines the accuracy of
$f\Delta t$, is inversely proportional to distance.   As a result, the
bound on $\lambda_g$ depends only on the measured masses of the objects and on
detector characteristics.  The only effect of distance is a weak $Z$-dependence 
arising from
cosmological effects.  
The results for the two kinds of detectors, and for various sources
are given in Table \ref{summary}.  
These correspond to bounds on the graviton rest
mass of order $2.5 \times 10^{-22}$ eV for ground-based, 
and $2.5 \times 10^{-26}$ eV for space-based observations.

Can bounds be placed on $\lambda_g$ using other observations or
experiments?  If the graviton is massive, then one expects that, in
the non-radiative near zone of a body like the Sun, the gravitational
potential will be modified from $GM /r$ to the Yukawa form 
\begin{equation}
V(r) = {GM \over r} \exp(-r/\lambda_g) \,.
\label{yukawa}
\end{equation}
Strictly speaking, such a conclusion would require a complete
gravitational theory of a massive graviton, capable of making
predictions both in the radiative and non-radiative regimes, and which
otherwise agrees with observation.  However, as several authors have
pointed out \cite{vandam,hare,nieto,visser}, construction of such a theory is a
non-trivial question.  Thus, in the absence of a well-defined
theoretical foundation, we shall make the phenomenological assumption
that, if the graviton is massive in the propagation of gravitational
waves, the Newtonian potential takes the form of Eq. (\ref{yukawa}), 
with the same
value of $\lambda_g$.

With this assumption, one can place bounds on $\lambda_g$ using solar-system 
dynamics.  Essentially, the orbits of the inner planets agree
with standard Newtonian gravity (including its post-Newtonian GR corrections)
to an accuracy of order $10^{-8}$.  Since the observed corrections to
Newtonian gravity in the limit $\lambda_g \gg r$ go as $(r/\lambda_g)^2$
(it is the acceleration, not the potential that is important),
this implies a rough bound $\lambda_g > 10^4$ astronomical units, or
$10^{12}\, {\rm km}$.   Talmadge {\it et al.} \cite{talmadge} surveyed solar
system data in the context of bounding the range and strength of 
a ``fifth force'', a Yukawa
term {\it added} to Newtonian gravity.  The best bound comes from
observations that verify Kepler's third law for the inner planets:
from observations of Mars, we find $\lambda_g > 2.8 \times 10^{12}$
km.  Bounds from other planets are summarized in Table
\ref{solarsystem}.
Apart from the Yukawa
potential assumption, this bound is solid and model independent.

Thus the bound inferred from gravitational radiation 
observations of stellar mass compact binary inspiral could be twice as
large as the solar-system bound, while that from supermassive binary
inspiral as observed by LISA could be $2 \times 10^4$ times larger.  

Some have argued for a larger bound on $\lambda_g$ from galactic and
cluster dynamics \cite{hiida,hare,nieto}, 
noting that the evidence of bound clusters
and of clear tidal
interactions between galaxies 
argues for a range $\lambda_g$ at least as
large as a few Megaparsecs ($6 \times 10^{19}$ km).  
Indeed this is the value quoted by the
Particle Data Group \cite{particle}.  However, in view of the uncertainties
related to the amount of dark matter in the universe, and the absence
of a theory that can encompass a massive graviton and cosmology, these
bounds should be viewed with caution.

The remainder of this paper provides the details underlying these
results.  In section II, we study the propagation of a massive
graviton in a cosmological background, to find the relation between
emission interval and arrival interval.  In section III, using the standard
``restricted PN approximation'', in which the gravitational waveform
is expressed as an amplitude accurate to the lowest, quadrupole
approximation, and a phase accurate through 1.5PN order [$O(v/c)^3$]
beyond the quadrupole approximation,
we determine the effect of graviton propagation time on the Fourier transform
of the waveform, which is the central ingredient in matched
filtering.  In section IV, we calculate the Fisher information matrix
and determine the accuracy with which the compact binary's parameters
can be measured, including a bound on the effect of graviton mass.
This approach is a reasonable approximation to real matched filtering
for Gaussian noise and large signal-to-noise ratio.   We apply the
results to specific noise curves and binary systems appropriate for
ground-based (LIGO/VIRGO) and space-based (LISA) detectors.
Section V discusses bounds on the graviton mass using solar-system
dynamics.  Henceforth, we use units in
which $G=c=1$.

\section{Propagation of a massive graviton}
\label{propagation}

Because some of the detectable compact binaries could be at
cosmological distances, we
study the propagation of a massive graviton in a background
Friedman-Robertson-Walker (FRW) homogeneous and isotropic spacetime.
We take the line element to have the form \cite{mtw}
\begin{equation}
ds^2=-dt^2+a^2(t)[d \chi^2 +\Sigma^2(\chi)(d\theta^2+\sin^2\theta
d\phi^2)] \,,
\label{cosmometric}
\end{equation}
where $a(t)$ is the scale factor of the universe and $\Sigma(\chi)$ is
equal to $\chi$, $\sin \chi$ or $\sinh \chi$ if the universe is
spatially flat, closed or open, respectively.  For a graviton moving
radially from emitter $\chi=\chi_e$ to detector $\chi=0$, it is
straightforward to show that the component of 4-momentum $p_\chi =$
constant.  Using the fact that $m_g^2=-p^\alpha p^\beta
g_{\alpha\beta}=E^2-a^{-2}p_\chi^2$, where $E=p^0$, 
together with $p^\chi/E=d\chi/dt$, we obtain
\begin{equation}
{d\chi \over dt} = - {1 \over a} \left ( 1+{m_g^2a^2 \over p_\chi^2}
\right )^{-1/2} \,,
\label{speed}
\end{equation}
where $p_\chi^2=a^2(t_e)(E_e^2-m_g^2)$.  Assuming that $E_e \gg m_g$,
expanding Eq. (\ref{speed}) to first order in $(m_g/E_e)^2$, and
integrating, we obtain
\begin{equation}
\chi_e= \int_{t_e}^{t_a} {dt \over a(t)} - {1 \over 2} {m_g^2 \over
{a^2(t_e)E_e^2}} \int_{t_e}^{t_a} a(t)dt \,.
\label{chi}
\end{equation}
Consider gravitons emitted at two different times $t_e$ and
$t_e^\prime$, with energies $E_e$ and $E_e^\prime$, and received at
corresponding arrival times ($\chi_e$ is the same for both).  
Assuming that $\Delta t_e \equiv
t_e-t_e^\prime \ll a/\dot a$, and noting that $m_g/E_e = (\lambda_g
f_e)^{-1}$, where $f_e$ is the emitted frequency, we obtain, after
eliminating $\chi_e$,
\begin{equation} 
\Delta t_a = (1+Z) \left [ \Delta t_e + {D \over 2\lambda_g^2} \left (
{1 \over f_e^2} -{1 \over {f_e^\prime}^2} \right ) \right ] \,,
\label{time}
\end{equation} 
where $Z \equiv a_0/a(t_e)-1$ is the cosmological redshift, and
\begin{equation} 
D \equiv {(1+Z) \over a_0}\int_{t_e}^{t_a} a(t) dt \,,
\label{D}
\end{equation} 
where $a_0=a(t_a)$ is the present value of the scale factor.
Note that $D$ is not a conventional cosmological distance measure,
like the luminosity distance $D_L \equiv a_0 \Sigma(\chi_e)(1+Z)$,
or the proper distance $D_P \equiv a_0\chi_e$.  For $Z \ll 1$, it
is given by the standard formula $D=Z/H_0$; for a matter dominated,
spatially flat universe, $D$ and $D_L$ are given by
\begin{mathletters}
\begin{eqnarray}
D &= & (2 / 5H_0) (1+Z)(1-(1+Z)^{-5/2}) \,. \label{2Da}\\
D_L &= & (2 / H_0) (1+Z)(1-(1+Z)^{-1/2}) \,. \label{2Db}
\end{eqnarray}
\label{2D} 
\end{mathletters}
The ratio $D/D_L$ will play a role in our analysis of the bound on
$\lambda_g$.  It has the following representative behavior:
\begin{equation}
{D \over D_L} = \cases{1-Z+O(Z^2) \,,&$Z \ll 1$, all $\Omega_0$\cr
  {{1+(2+Z)(1+Z+\sqrt{1+Z})} \over {5(1+Z)^2}} \,,&$\Omega_0=1$, all
$Z$}
\label{DbyDL}
\end{equation}
where $\Omega_0$ is the density parameter.
At $Z=1$, the factor $D/D_L$ varies from 0.5 for $\Omega_0=0.01$
to 0.6 for $\Omega_0=2$.  For simplicity, we shall henceforth assume that
$\Omega_0 \equiv 1$.

\section{Massive graviton propagation and the phasing of gravitational
waves}

We shall treat the problem of a binary system of compact bodies of
locally measured masses $m_1$ and $m_2$ in a
quasi-circular orbit, that is an orbit which is circular apart from an
adiabatic inspiral induced by gravitational radiation reaction within
GR.  We ignore tidal interactions and spin effects.
For matched filtering of gravitational waves using LIGO/VIRGO or LISA
type detectors, it is sufficient 
for our purpose to write the gravitational waveform $h(t)$
in the ``restricted post-Newtonian form'' \cite{cutlerflan,finn,jugger}, 
in terms of an
amplitude $A(t)$ expressed to the lowest, quadrupole approximation,
and a phase $\Phi(t)$, expressed as a post-Newtonian expansion several
orders beyond the quadrupole approximation, 
\begin{mathletters}
\begin{eqnarray}
h(t) &\equiv& A(t)e^{-i\Phi(t)} \,,
\label{hbasic}\\
\Phi(t) &\equiv& \Phi_c + 2\pi \int_{t_c}^t f(t) dt \,,
\label{Phi}
\end{eqnarray}
\end{mathletters}
where $f(t)$ is the observed frequency of the waves, and $\Phi_c$ and
$t_c$ are ``fiducial'' phase and time respectively.  The amplitude $A$
is given by
\begin{equation}
A(t) = {2\mu \over {a_0 \Sigma(\chi_e)}} 
{m \over r(t)}F(i,\theta,\phi,\psi) \,, 
\label{amplitude}
\end{equation}
where $m \equiv m_1+m_2$ and $\mu \equiv m_1m_2/m$ are 
the total and reduced mass of
the system (we also define the reduced mass parameter $\eta \equiv \mu/m$),
$r(t)$ is the orbital separation, and $F$ is an angular function
related to the orientation of the orbit (angles $i$, $\psi$) 
and the direction of the source relative to the antenna
(angles $\theta$, $\phi$), given by
\begin{equation}
F^2(i,\theta,\phi,\psi)= {1 \over 4}(1+\cos^2 i)^2 F_+^2 +\cos^2 i
F_\times^2 \,,
\label{F}
\end{equation}
where $F_+(\theta,\phi,\psi)$ and $F_\times(\theta,\phi,\psi)$ are beam
pattern factors quoted, for example in Eqs. (104) of \cite{Thorne1987}.
For simplicity, we shall average over all four angles, and use the
fact that $\langle F^2 \rangle = 4/25$.   

We next compute the Fourier transform of $h(t)$.  Expanding $h(t)$
about the time $\tilde t$ at which the observed frequency is $\tilde f$, {\it
i.e.} $f(\tilde t)\equiv \tilde f$, and using the stationary-phase
approximation, we obtain
\begin{equation}
\tilde h (\tilde f) = {A(\tilde t) \over \sqrt{\dot f (\tilde t)}}
e^{i\Psi(\tilde f)} \,,
\label{htilde}
\end{equation}
where
\begin{mathletters}
\begin{eqnarray}
A(\tilde t) &=& {4 \over 5}{{\cal M}_e 
\over {a_0 \Sigma(\chi_e)}} (\pi{\cal M}_e
\tilde f_e)^{2/3} \,,\label{A}\\
\Psi(\tilde f) &=& 2\pi\int_{f_c}^{\tilde f} (t-t_c)df +2\pi \tilde f t_c
-\Phi_c-\pi/4 \,,
\label{Psi}
\end{eqnarray}
\end{mathletters}
where ${\cal M}_e=\eta^{3/5}m$ is the ``chirp'' mass of the emitter, 
and where we have
used the Newtonian relation $m/r(\tilde t) = (\pi m \tilde f_e)^{2/3}$.
The subscript
``e'' denotes ``at the emitter''.  
We next substitute Eq. (\ref{time}) into (\ref{Psi}) to relate the
time at the detector to that at the emitter, noting that, because of
the cosmological redshift, $f_e = (1+Z)f$. The result is 
\begin{equation}
\Psi(\tilde f) = 2\pi\int_{\tilde f_{ec}}^{\tilde f_e} (t_e-t_{ec})df_e 
-{{\pi D} \over {f_e \lambda_g^2}}+2\pi \tilde f \bar t_c - \bar
\Phi_c - {\pi \over 4} \,,
\label{Psif}
\end{equation}
where $\bar t_c=t_c-D/[2(1+Z)\lambda_g^2 f_c^2]$, and 
$\bar \Phi_c=\Phi_c-2\pi D/[(1+Z)\lambda_g^2 f_c]$.
To find $t_e-t_{ec}$ as a function
of $f_e$, we integrate the equation for radiation reaction between
$t_{ec}$ and $t_e$:
\begin{eqnarray}
{df_e \over dt_e} = {96 \over {5\pi {\cal M}_e^2}} (\pi {\cal M}_ef_e)^{11/3}
  && \biggl [ 1- \left ( {743 \over 336}+{11 \over 4}\eta \right ) (\pi m
f_e)^{2/3} \nonumber \\   
&&+ 4\pi(\pi mf_e) \biggr ] \,,
\label{fdot}
\end{eqnarray}
where we have included the first post-Newtonian (PN) term and the
1.5PN ``tail'' term in the radiation-reaction equation (see, {\it
e.g.} \cite{cutlerflan}).  After
absorbing further constants of integration
into $\bar t_c$ and $\bar \Phi_c$, dropping the bars on those two
quantities, and re-expressing everything in
terms of the {\it measured} frequency $\tilde f$ [note that 
$(\dot f)^{1/2}=(df_e/dt_e)^{1/2}/(1+Z)$], we obtain
\begin{mathletters}
\begin{eqnarray}
\tilde h (\tilde f) &=& \cases{ \tilde A(\tilde f) 
e^{i\Psi(\tilde f)} \,,& $0<\tilde f< \tilde f_{\rm max} $\cr
0 \,,& $\tilde f> \tilde f_{\rm max} $} \label{tildehnew}\\
A(\tilde f) &\equiv&
 {\cal A} \tilde f^{-7/6} = \sqrt{\pi \over 30}{{\cal M}^2 \over D_L} u^{-7/6} 
\,, \label{Anew}\\
\Psi(\tilde f) &=& 2\pi \tilde f t_c
-\Phi_c-\pi/4 +{3 \over 128} u^{-5/3} - \beta u^{-1} \nonumber\\
&&+{5 \over 96} \left ( {743\over 336}+{11\over 4}\eta \right )
\eta^{-2/5} u^{-1} \nonumber \\
&&- {3\pi \over 8} \eta^{-3/5} u^{-2/3} \,,
\label{Psinew}
\end{eqnarray}
\label{fourierfinal}
\end{mathletters}
where $u \equiv \pi {\cal M} \tilde f$, and $\cal M$ is the ``measured
chirp mass'', related to the source chirp mass by a redshift: ${\cal M} =
(1+Z) {\cal M}_e$.  The parameter $\beta$ is given by
\begin{equation}
\beta \equiv {{\pi^2 D {\cal M}} \over {\lambda_g^2 (1+Z)}} \,.
\label{beta}
\end{equation}
The frequency $\tilde f_{\rm max}$ represents an upper cut-off
frequency where the PN approximation fails.
Equations (\ref{fourierfinal}) are the basis for an analysis of
parameter estimation using matched filtering.

Before turning to matched filtering, we must address our approximation
of the motion and gravitational radiation damping as being general relativistic
up to corrections of order $(r/\lambda_g)^2$.  In the radiation-reaction formula
Eq. (\ref{fdot}),
we included corrections to the quadrupole formula at 1.5PN order,
corresponding to corrections of order $v^3$.  Thus our neglect of
massive graviton effects amounts to assuming that $r^2 \lambda_g^{-2} v^{-3}
\ll 1$ for all systems of interest.  Because $v^2 \simeq m/r$ for
circular orbits, we can rewrite this condition as
$(m/\lambda_g)v^{-5/2} \ll 1$.  Since typically $v > 10^{-2}$ for all
systems of interest, and $\lambda_g > 10^{12} \,{\rm km}$ from
solar-system bounds,
this condition is easily satisfied.

\section{Bounds on the graviton mass using matched filtering}

\subsection{Matched-filter analysis}

To obtain a more reliable 
estimate of the bound that can be placed on the graviton
mass, 
we carry out a full matched-filter analysis following the 
method outlined for compact binary inspiral by Cutler and 
Flanagan \cite{cutlerflan} and Finn and Chernoff \cite{finn}.  
The details here parallel
those of \cite{pnfilters}.

With a given noise spectrum $S_n(f)$, one defines the inner product of
signals $h_1$ and $h_2$ by 
\begin{equation}
(h_1|h_2) \equiv 2\int_0^\infty {{\tilde h_1^* \tilde h_2 +\tilde
h_2^* \tilde h_1 } \over S_n(f)} df \,,
\label{innerproduct}
\end{equation}
where $\tilde h_a$ is the Fourier transform of the waveform defined in
Eqs. (\ref{fourierfinal}) (henceforth, we drop the tilde on
frequencies).  The signal-to-noise ratio for a given
signal $h$ is given by 
\begin{equation}
\rho[h] \equiv S/N[h] = (h|h)^{1/2} \,.
\label{signaltonoise}
\end{equation}
If the signal depends on a set of parameters $\theta^a$ which are to
be estimated by the matched filter, then the rms error in $\theta^a$
in the limit of large $S/N$ is given by 
\begin{equation}
\Delta \theta^a \equiv \sqrt{\langle (\theta^a - \langle \theta^a
\rangle )^2 \rangle } = \sqrt{ \Sigma^{aa}} \,,
\label{deltatheta}
\end{equation}
where $\Sigma^{aa}$ is the corresponding component of the inverse of
the covariance matrix or Fisher information matrix $\Gamma_{ab}$
defined by 
\begin{equation}
\Gamma_{ab} \equiv \left ( {{\partial h} \over {\partial \theta^a}}
\big | {{\partial h} \over {\partial \theta^b}} \right ) \,.
\label{fisher}
\end{equation}
The correlation coefficient between two parameters $\theta^a$ and
$\theta^b$ is 
\begin{equation}
c^{ab} \equiv \Sigma^{ab}/\sqrt{\Sigma^{aa}\Sigma^{bb}} \,.
\label{correlation} 
\end{equation}

We estimate the following six parameters, $\ln {\cal A}$, $\Phi_c$, $f_0
t_c$, $\ln {\cal M}$, $\ln \eta$, and $\beta$, where $f_0$ is a frequency
characteristic of the detector, typically a ``knee'' frequency, or a
frequency at which $S_n(f)$ is a minimum.  The corresponding partial
derivatives of $\tilde h (f)$ are
\begin{mathletters}
\begin{eqnarray}
{{\partial \tilde h(f)} \over {\partial \ln {\cal A} }} &=& \tilde
h(f)  
\,, \\
{{\partial \tilde h(f)} \over {\partial \Phi_c }} &=& -i\tilde
h(f)
\,, \\
{{\partial \tilde h(f)} \over {\partial f_0t_c }} &=& 2\pi i (f/f_0) \tilde
h(f) 
\,, \\
{{\partial \tilde h(f)} \over {\partial \ln {\cal M} }} &=& -\left (
  {5i \over 128} u^{-5/3} + {5i \over 96} \gamma(\eta) u^{-1} \right .
\nonumber \\
  && \left .  -{i\pi
\over 4} \eta^{-3/5} u^{-2/3} \right )\tilde h(f)
\,, \\
{{\partial \tilde h(f)} \over {\partial  \ln\eta }} &=&   \left (
  {5i \over 96} \eta \gamma^\prime (\eta) u^{-1} \right . \nonumber
\\
&& \left . +{9i\pi \over 40}
\eta^{-3/5} u^{-2/3} \right ) \tilde
h(f)
\,, \\
{{\partial \tilde h(f)} \over {\partial \beta }} &=&   -i u^{-1} \tilde
h(f)
\,,
\end{eqnarray}
\label{partials}
\end{mathletters}
where $\gamma(\eta) \equiv (743/336 + 11\eta/4)\eta^{-2/5}$, and
$\gamma^\prime \equiv d\gamma/d\eta$.  Since we plan to derive the
error in estimating $\beta$ about the nominal or {\it a priori} GR
value $\beta=0$, we have set $\beta=0$ in all the partial derivatives.

We assume that the detector noise curve can be approximated by an
amplitude $S_0$, which sets the overall scale of the noise, and a
function of the ratio $f/f_0 \equiv x$, which may include additional
parameters, that is $S_n(f) = S_0 g_\alpha (x)$, where the subscript $\alpha$
denotes a set of parameters.  Then from Eqs.
(\ref{fourierfinal}) and (\ref{signaltonoise}) we find that the
signal-to-noise ratio is given by  
\begin{eqnarray}
\rho &=& 2 {\cal A} f_0^{-2/3} (I(7)/S_0)^{1/2} \nonumber \\
&=& \sqrt{2 \over 15} {{\cal M}^{5/6} \over D_L} (\pi
f_0)^{-2/3} \left ( {I(7) \over S_0} \right )^{1/2} \,,
\label{SN}
\end{eqnarray}
where  we define the integrals
\begin{equation}
I(q) \equiv \int_0^\infty { x^{-q/3} \over g_\alpha(x)} dx \,.
\label{Iq}
\end{equation}
Note that any frequency cut-offs are to be incorporated appropriately
into the endpoints of the
integrals $I(q)$.  If we define the coefficients $I_q \equiv
I(q)/I(7)$, then all elements of the covariance
matrix turn out to be proportional to $\rho^2$ times linear
combinations of terms of the form $u_0^{-n/3}I_q$ for various integers
$n$ and $q$, where $u_0 = \pi {\cal M} f_0$.  This overall $\rho$
dependence is characteristic of the large $S/N$ limit.
As a result, the rms errors $\Delta \theta^a$ are inversely
proportional to $\rho$, while the correlation coefficients are
independent of $\rho$.  Defining $\Delta \beta \equiv \Delta^{1/2} /\rho$,
viewing $\Delta \beta$ as an upper bound on $\beta$, and combining
this definition with Eqs. (\ref{beta}) and (\ref{SN}) we obtain the
{\it lower} bound on $\lambda_g$:
\begin{equation}
\lambda_g >  \left ( {2 \over 15}{I(7) \over S_0} \right )^{1/4} \left ( {D \over (1+Z)D_L} \right )^{1/2} {{\pi^{2/3} {\cal M}^{11/12}} \over
{f_0^{1/3} \Delta^{1/4}}} \,.
\label{lambdabound}
\end{equation}
Note that the bound on $\lambda_g$ depends only weakly on distance, via
the $Z$ dependence of the factor $[D/(1+Z)D_L]^{1/2}$, which varies
from unity at $Z=0$ to 0.45 at $Z=1.5$.  This is because, while
the signal strength, and hence the accuracy, falls with distance, the
size of the arrival-time effect increases with distance.  Otherwise,
the bound depends only on the chirp mass and on detector
noise characteristics.  We now apply this formalism to specific
detectors.

\begin{table}[t]
\twocolumn[\hsize\textwidth\columnwidth\hsize\csname
           @twocolumnfalse\endcsname
\caption{The rms errors for signal parameters, the corresponding
bound on $\lambda_g$, in units of $10^{12}$ km, and the correlation coefficients
$c_{{\cal M}\eta}$, $c_{{\cal M} \beta}$ and $c_{ \beta \eta}$.
The noise
spectrum is that of the advanced LIGO system, given by
Eq. (\protect\ref{ligonoise}), and a signal-to-noise ratio of 10
is assumed.  Masses are in units of $M_\odot$, $\Delta t_c$ is
in msec.  }
\begin{tabular}{cccccccccc}
$m_1$&$m_2$&$\Delta \phi_c$&$\Delta t_c$&$\Delta {\cal M}/{\cal M}$
&$\Delta \eta/\eta$&$\lambda_g$&$c_{{\cal M}\eta}$&$c_{{\cal M}
\beta}$&$c_{\beta \eta}$\\
\hline
1.4&1.4&4.09&1.13&0.034\%&7.88\%&4.6&-0.971&-0.993&0.992\\
1.4&10.0&6.24&2.04&0.191\%&12.2\%&5.4&-0.978&-0.994&0.994\\
10.0&10.0&9.26&3.53&1.42 \%&57.3\%&6.0&-0.983&-0.994&0.997\\
\end{tabular}
\label{tableLIGO}
\vskip 2pc]
\end{table}

\narrowtext

\subsection{Ground-based detectors of the LIGO/VIRGO type}

The proposed advanced version of LIGO is expected to detect
compact binary inspiral to distances of 200 Mpc to 1 Gpc.  The
sensitive frequency band extends from around 10 Hz to several hundreds
of Hz.  We adopt the benchmark advanced LIGO noise curve, given by
\begin{equation}
S_n(f) = \cases{\infty \,,&$f< 10 {\rm Hz} $\cr
   S_0 [(f_0/f)^4+2+2(f/f_0)^2]/5 \,,&$f> 10{\rm Hz} $}
\label{ligonoise}
\end{equation}
where $S_0=3 \times 10^{-48} \,{\rm Hz}^{-1}$, and $f_0=70 \,{\rm
Hz}$.  The cutoff at 10 Hz corresponds to seismic noise, while the
$f^{-4}$ and $f^2$ dependences denote thermal and photon shot-noise,
respectively \cite{LIGO}.  We choose an upper cut-off frequency, where the PN
approximation fails, corresponding to the innermost stable circular 
orbit.  Although this is known rigorously only for test body motion
around black holes \cite{kww}, a conventional estimate is given by $f_{\rm
ISCO} \approx [6^{3/2} \pi (m_1+m_2)]^{-1}$.  Converting this to the
measured frequency and chirp mass, 
we have $x_{\rm max} = [6^{3/2}\pi \eta^{-3/5}
{\cal M} f_0]^{-1}$.  For this case, we thus have $g(x)=(x^{-4}+2+2x^2)/5$, and
$I(q)=\int_{1/7}^{x_{\rm max}} [x^{-q/3}/g(x)]dx$.  We then calculate and
invert the covariance matrix 
and evaluate the errors in the five relevant parameters (the parameter
$\ln {\cal A}$ decouples from the rest and is relevant only for the
calculation of $\rho$), 
and the correlation coefficients between $\cal M$, $\eta$
and $\beta$.  For various ``canonical'' compact binary systems
observable by advanced LIGO, the results are shown in Table \ref{tableLIGO}.  
Note that, in determining the bound on $\lambda_g$, we must include
the $Z$ dependence embodied in Eq. (\ref{lambdabound}).  To do so, we take
our assumed value for signal-to-noise ratio $\rho=10$, determine the
luminosity distance using Eq. (\ref{SN}), and convert that to a redshift
using Eq. (\ref{2Db}), with an assumed value $H_0=50 \, {\rm km~s}^{-1}{\rm
Mpc}^{-1}$ and $\Omega_0=1$.  We then substitute it, along with 
Eq. (\ref{2Da}) into Eq.
(\ref{lambdabound}).  

It is useful to compare these results to
those from parameter estimation calculations
using
pure GR to 1.5PN order {\it including} spin-orbit effects
(see {\it e.g.} \cite{cutlerflan,pnfilters}).  There, an additional
parameter related to the spin-orbit effect (also called $\beta$, with
a nominal value of zero) was
estimated, although it produced a different $u$-dependent term in the
phasing formula ($u^{-2/3}$ instead of $u^{-1}$).  Nevertheless, the
errors in the fiducial phase $\Delta \Phi_c$, time
$\Delta t_c$ and
chirp mass $\Delta \ln {\cal M}$ are virtually identical in both cases, and
somewhat larger than if no additional $\beta$ parameter were
estimated (compare Table \ref{tableLIGO} with Table I and II of
\cite{cutlerflan} or Table II of \cite{pnfilters}).
But in our 
case, the errors
in the reduced mass parameter $\eta$ are significantly larger, a
result of the nearly perfect correlation ($u^{-1}$ dependence) 
between the 1PN term and the
$\beta$-term in the phasing, Eq. (\ref{Psinew}).  The error grows
dramatically with total mass because the smaller number of observed
gravitational-wave cycles reduces the ability of the tail term
($\propto u^{-2/3}$) to break the degeneracy.  

\begin{table}[t]
\twocolumn[\hsize\textwidth\columnwidth\hsize\csname
           @twocolumnfalse\endcsname
\caption{Rms errors on signal parameters, the bound on $\lambda_g$, in
units of $10^{16}$ km, and the correlation coefficients.
The noise spectrum is that of
LISA including white-dwarf binary confusion noise, given by Eq.
(\protect\ref{lisanoise}).  Signal-to-noise
ratio $\rho$ is shown, corresponding to a luminosity distance of about
3 Gpc. Masses are in units of $M_\odot$, $\Delta t_c$ is
in sec. }
\begin{tabular}{ccccccccccc}
$m_1$&$m_2$&$\rho$&$\Delta \phi_c$&$\Delta t_c$&$\Delta {\cal M}/{\cal M}$
&$\Delta \eta/\eta$&$\lambda_g$&$c_{{\cal M}\eta}$&$c_{{\cal M}
\beta}$&$c_{\beta \eta}$\\
\hline
$10^7$&$10^7$&1600&0.073&20.0&0.0187\%&0.562\%&6.9&-0.979&-0.992&0.997\\
$10^7$&$10^6$&710&0.145&22.5&0.0119\%&0.362\%&3.9&-0.984&-0.995&0.997\\
$10^6$&$10^6$&5800&0.017&0.48&0.0021\%&0.108\%&5.4&-0.954&-0.985&0.991\\
$10^6$&$10^5$&4300&0.026&0.40&0.0015\%&0.062\%&3.0&-0.970&-0.992&0.991\\
$10^5$&$10^5$&2100&0.017&0.09&0.0008\%&0.072\%&2.3&-0.946&-0.975&0.992\\
$10^5$&$10^4$&750&0.048&0.18&0.0007\%&0.059\%&1.2&-0.955&-0.987&0.989\\
$10^4$&$10^4$&320&0.092&0.22&0.0004\%&0.141\%&0.7&-0.963&-0.992&0.989\\
\end{tabular}
\label{tableLISA}
\vskip 2pc]
\end{table}

\subsection{Space-based detectors of the LISA type}

The proposed Laser Interferometer Space Antenna (LISA) is expected to
be able to detect the inspiral of supermassive black hole binaries to
cosmological distances, with very large signal-to-noise ratio.  The
sensitive frequency band extends from around $10^{-4}$ to $10^{-1}$
Hz, with a typical integration time of the order of one year.
We adopt a noise curve described in the LISA pre-Phase A report
\cite{LISA},
augmented by a fit to ``confusion noise'' generated by a population of
close white dwarf binaries in our galaxy \cite{bender}, given by
the equations:
\begin{eqnarray}
S_0 &=& 4.2 \times 10^{-41} {\rm Hz}^{-1} \,, \nonumber \\
f_0 &=& 10^{-3} {\rm Hz} \,, \nonumber \\
g_(x) &=& \sqrt{10} x^{-14/3} +1+ x^2/1000 \nonumber \\
&&+ 313.5 x^{-(6.398+3.518
\log_{10} x )} \,.
\label{lisanoise}
\end{eqnarray}
In order, the four terms in $g(x)$ correspond to: temperature
fluctuations, photon shot noise, loss of sensitivity when the arm
lengths exceed the gravitational wavelength, and a fit to the 
white-dwarf binary
confusion noise.  For the maximum frequency, we again adopt that of
the innermost stable circular orbit.  The minimum frequency is set by
the characteristic integration time for LISA, nominally chosen to be
one year.  We calculate the
time $T_e$ remaining until the system reaches
the innermost stable orbit by integrating Eq.
(\ref{fdot}) using only the dominant, Newtonian contribution, convert
from time at the emitter to observation time $T$ using Eq. (\ref{time})
(ignoring the small correction due to
massive graviton propagation), and obtain
\begin{equation}
x_{\rm min} \approx {1 \over {\pi {\cal M} f_0}} \left ( {5{\cal M}
\over 256T} \right )^{3/8} \,.
\label{lisaxmin}
\end{equation}

For supermassive binaries ranging from $10^4$ to $10^7 M_\odot$, and
for integration time $T=$ 1 year, we estimate the errors in the five
parameters and determine a bound on $\lambda_g$.  We choose the
signal-to-noise ratio $\rho$ for each case such that the luminosity
distance to the source $\simeq 3$ Gpc, so that cosmological effects do
not become too severe.  The results are shown in Table
\ref{tableLISA}.

\section{Solar-system bounds on the graviton mass}

If the Newtonian gravitational potential is modified by a massive
graviton to have the Yukawa form of Eq. (\ref{yukawa}), then the
acceleration of a test body takes the form
\begin{equation}
{\bf g} = -{{\bf n} \over r^2} \mu(r) \,,
\label{acceleration}
\end{equation}
where 
\begin{eqnarray}
\mu(r) &\equiv& M(1+r/\lambda_g)\exp (-r/\lambda_g) \nonumber \\
&=& M \left [ 1 - {1 \over 2} \left ({r \over \lambda_g} \right )^2 
+O\left ({r \over \lambda_g} \right )^3 \right ] \,.
\label{kepler}
\end{eqnarray}
~For a planet with semi-major axis $a_p$ and period $P_p$, Kepler's
third law gives $a_p (2\pi/P_p)^{2/3} = \mu(a_p)^{1/3}$.  For a pure
inverse-square law, $\mu \equiv {\rm constant}$, and its value is
determined accurately using the orbit of the Earth.  Thus, by checking Kepler's
third law for other planets, one can test the constancy of $\mu$.
For a given planet, we define the parameter $\eta_p$ by 
\begin{equation}
1+\eta_p \equiv \left ({\mu(a_p) \over \mu(a_\oplus)} \right )^{1/3} \,.
\label{eta}
\end{equation}
Combining Eq. (\ref{eta}) and (\ref{kepler}), we obtain a bound on $\lambda_g$
in terms of $\eta_p$
\begin{equation}
\lambda_g > \left ( {1-a_p^2} \over {6\eta_p} \right )^{1/2} \,,
\label{lambdasolar}
\end{equation}
where $\lambda_g$ and $a_p$ are expressed in astronomical units ($1.5 \times
10^8$ km).  Table \ref{solarsystem} lists the observed bounds on
$\eta_p$ for Mercury, Venus, Mars and Jupiter compiled by Talmadge {\it
et al.} \cite{talmadge}, and the resulting bounds on $\lambda_g$.  

\begin{table}[t]
\caption{Bounds on $\lambda_g$ in units of $10^{12}$ km 
from Kepler's third law applied to the
solar system.   Semi-major axes are in astronomical units, and the
appropriate one-sided, $2\sigma$ bound on $\eta_p$ from Talmadge {\it
et al.} \protect\cite{talmadge}
is shown.}
\begin{tabular}{lccc}
Planet&$a_p$&Bound on $\eta_p$&$\lambda_g$\\
\hline
Mercury&0.387&$1.4 \times 10^{-8}$&0.5\\
Venus&0.723&$1.5 \times 10^{-9}$&1.1\\
Mars&1.523&$-6.5 \times 10^{-10}$&2.8\\
Jupiter&5.203&$-6 \times 10^{-8}$&1.3\\
\end{tabular}
\label{solarsystem}
\end{table}

A Yukawa violation of the inverse square law will also produce
anomalous perhelion shifts of orbits, of the form $\delta \omega
\approx \pi (a_p/\lambda_g)^2$.  By comparing 
measured shifts with the general relativistic prediction for Mercury
and Mars (Venus and Jupiter are unsuitable for this purpose) 
\cite{talmadge}, one
obtains bounds a factor two weaker than those from Kepler's law.

Systems like the binary pulsar do not yield useful bounds, mainly
because they are so compact.  Since the deviations from GR go as
$(a/\lambda_g)^2$, where $a$ is the size of the system, the bound is
roughly $\lambda_g > a/\epsilon^{1/2}$, where $\epsilon$ is the
accuracy of agreement with observations.  Since $a \sim 10^6$ km, and
$\epsilon \sim 10^{-3}$, the bound is far from competitive with that
from the solar system.

\section*{Acknowledgments}

We are grateful to the Racah Institute of Physics, Hebrew University
of Jerusalem, where part of this work was carried out.  This work was
supported in part by the J. S. Guggenheim Foundation, and by the
National Science Foundation Grant No. PHY 96-00049.

\end{document}